\documentclass[pra,twocolumn,superscriptaddress,groupedaddress]{revtex4}    
\usepackage{graphicx}
\usepackage{epstopdf}
\usepackage{amssymb}
\usepackage{color}
\usepackage{amsfonts}
\usepackage{subfigure}
\usepackage{bm}
\usepackage{epstopdf}

\begin{document}



\title{De-excitation spectroscopy of strongly interacting Rydberg gases}

\author{C. Simonelli}
\affiliation{Dipartimento di Fisica ``E. Fermi'', Universit\`a di Pisa, Largo Bruno Pontecorvo 3, 56127 Pisa, Italy}
\affiliation{INO-CNR, Via G. Moruzzi 1, 56124 Pisa, Italy}

\author{M. Archimi}
\affiliation{Dipartimento di Fisica ``E. Fermi'', Universit\`a di Pisa, Largo Bruno Pontecorvo 3, 56127 Pisa, Italy}

\author{L. Asteria}
\affiliation{Dipartimento di Fisica ``E. Fermi'', Universit\`a di Pisa, Largo Bruno Pontecorvo 3, 56127 Pisa, Italy}

\author{D. Capecchi}
\affiliation{Dipartimento di Fisica ``E. Fermi'', Universit\`a di Pisa, Largo Bruno Pontecorvo 3, 56127 Pisa, Italy}

\author{G. Masella}
\affiliation{Dipartimento di Fisica ``E. Fermi'', Universit\`a di Pisa, Largo Bruno Pontecorvo 3, 56127 Pisa, Italy}
\affiliation{INO-CNR, Via G. Moruzzi 1, 56124 Pisa, Italy}

\author{E. Arimondo}
\affiliation{Dipartimento di Fisica ``E. Fermi'', Universit\`a di Pisa, Largo Bruno Pontecorvo 3, 56127 Pisa, Italy}
\affiliation{INO-CNR, Via G. Moruzzi 1, 56124 Pisa, Italy}
\affiliation{CNISM UdR Dipartimento di Fisica ``E. Fermi'', Universit\`a di Pisa, Largo Pontecorvo 3, 56127 Pisa, Italy}

\author{D. Ciampini}
\affiliation{Dipartimento di Fisica ``E. Fermi'', Universit\`a di Pisa, Largo Bruno Pontecorvo 3, 56127 Pisa, Italy}
\affiliation{INO-CNR, Via G. Moruzzi 1, 56124 Pisa, Italy}
\affiliation{CNISM UdR Dipartimento di Fisica ``E. Fermi'', Universit\`a di Pisa, Largo Pontecorvo 3, 56127 Pisa, Italy}

\author{O. Morsch}
\affiliation{Dipartimento di Fisica ``E. Fermi'', Universit\`a di Pisa, Largo Bruno Pontecorvo 3, 56127 Pisa, Italy}
\affiliation{INO-CNR, Via G. Moruzzi 1, 56124 Pisa, Italy}




\date{\today}

\begin{abstract}
We present experimental results on the controlled de-excitation of Rydberg states in a cold gas of $Rb$ atoms. The effect of the van der Waals interactions between the Rydberg atoms is clearly seen in the de-excitation spectrum and dynamics. Our observations are confirmed by numerical simulations. In particular, for off-resonant (facilitated) excitation we find that the de-excitation spectrum reflects the spatial arrangement of the atoms in the quasi one-dimensional geometry of our experiment. We discuss future applications of this technique and implications for detection and controlled dissipation schemes.
\end{abstract}

\maketitle

Cold atoms excited to high-lying Rydberg states have become a thriving field of research in recent years. Their strong and widely tunable interactions make them an attractive and versatile platform for studying a range of many-body phenomena such as collective excitations  \cite{Ga�tan:2009, G�rttner:2014}  and spatial ordering \cite{Schau�:2012}, and for classical and quantum simulations of, e.g., glassy systems and percolation \cite{Urban:2009, Saffman:2013, Lesanovsky:2013, Urvoy:2015,Valado:2016, Marcuzzi:2016}. Typically, in such experiments Rydberg states are excited starting from a cold or Bose-condensed cloud of ground-state atoms, and the ensuing dynamics is then studied using field ionization techniques or controlled depumping to the ground state \cite{Schau�:2012, Dudin:2012}. Whereas the number of ground-state atoms is usually on the order of tens or hundreds of thousands, only a few to tens of Rydberg excitations are created in most experiments. The Rydberg blockade, whereby an excited atom suppresses further excitations within a volume defined by a blockade radius $r_b$, and the facilitation mechanism, which favours excitations inside a shell at a well-defined distance $r_{fac}$ in the case of off-resonant excitation, both restrict the number of excitations possible in an atomic cloud to a small fraction of the total number of atoms contained in it. Nevertheless, in order to fully describe the excitation dynamics, theoretical treatments and numerical simulations need to take into account all the atoms in the cloud, particularly so in the coherent excitation regime, where the van der Waals interaction between Rydberg atoms leads to collective effects. 
 
Here we present experimental results and numerical simulations of the reverse process, i.e., a controlled de-excitation of a cold Rydberg gas induced by resonantly coupling the Rydberg state to a short-lived intermediate state that decays to the ground state. We perform experiments in different regimes, ranging from non-interacting to strongly interacting, and we show that since only excited atoms are involved, the dynamics of the de-excitation process allows a much simpler description and permits studies of interaction effects that are complementary to the conventional experiments starting from ground-state atoms. Our method can be used as a sensitive tool for probing the spatio-temporal evolution of an interacting Rydberg gas. We also point out possible consequences for depumping-based detection protocols and for controlled dissipation schemes. 

In our experiments we create small clouds of cold $^{87} Rb$ atoms in a magneto-optical trap (MOT). The clouds have approximately Gaussian density profiles with typical widths around $150\,\mathrm{\mu m}$ and peak densities of $10^{11} \,\mathrm{cm^{-3} }$. Two laser beams at $420$ and $1013$ nm, respectively, couple the $5S_{1/2}$ ground state to the $70S_{1/2}$ Rydberg state through a coherent two-photon excitation with Rabi frequency $\Omega=\Omega_{1013}\Omega_{420}/2\Delta_{6P} $ via the intermediate $6P_{3/2}$ state, from which the lasers are detuned by $\Delta_{6P}\approx 2\pi \times 40\,\mathrm{MHz}$ (in this work,  $\Delta_{6P}$ is limited by the frequencies of the acousto-optic modulator (AOM) used to switch between excitation and de-excitation, see below). Here, $\Omega_{420}$ and $\Omega_{1013}$ are the Rabi frequencies of the two transitions. The laser at $420\,\mathrm{nm}$ is focused to a $6 \,\mathrm{\mu m}$ waist, while the $1013 \,\mathrm{nm}$  laser has a larger waist of around $130 \,\mathrm{\mu m}$. The resulting interaction volume is defined by the overlap of the two quasi co-propagating laser beams \cite{Simonelli:2016}. As the typical distances between Rydberg excitations in our experiment, the blockade radius $r_b \approx 10 \,\mathrm{\mu m}$ and the facilitation radius $r_{fac} \approx 6 \,\mathrm{\mu m}$ \cite{Lesanovsky:2014,Valado:2016}, are comparable to the radial size of the interaction volume, the probability of exciting more than one atom in the radial direction is strongly suppressed. The excitation dynamics can, therefore, be considered to be quasi one-dimensional. Furthermore, while our excitation process is, in principle, coherent, in practice the total transition linewidth $\gamma= 2 \pi \times 700 \,\mathrm{kHz}$ (due to the individual laser linewidths and residual atomic motion for our MOT at $T \approx 150\,\mathrm{\mu K}$) limits the coherence time to 1-2  $\,\mathrm{\mu s}$. In most of the results reported here, the excitation process can, therefore, be described by an excitation rate $\Gamma$ that depends both on the detuning $\Delta_{ex}$ from resonance and the interactions with all the other Rydberg atoms, as reported in \cite{Lesanovsky:2013}. In the single atom regime  the excitation rate is expressed by the formula $\Gamma= \frac{\Omega^2}{2 \gamma}\cdot\frac{1}{|1+(\Delta_{ex}/\gamma)^2|}$ and can be easily extended to the interacting regime by adding the interaction energy contribution to the detuning $\Delta_{ex}$.

\begin{figure}[htbp]
\begin{center}
\includegraphics[width=4cm]{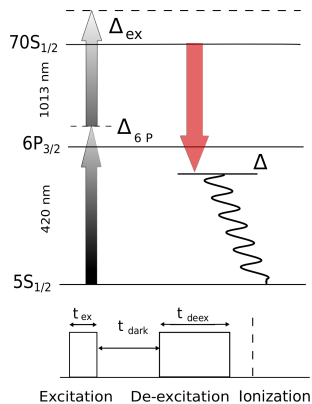}
\caption{Excitation / de-excitation protocol. First, atoms are excited to the $70S_{1/2}$ Rydberg state (with a detuning $\Delta_{ex}$) by two laser beams at $420\,\mathrm{nm}$ and $1013 \,\mathrm{nm}$, respectively. The  $420\,\mathrm{nm}$ laser is detuned by $\Delta_{6P} = 2\pi \times 40  \,\mathrm{MHz}$ from the intermediate $6P_{3/2}$ state.  After a variable dark time $t_{dark}$, the frequency of the laser at $1013 \,\mathrm{nm}$ is shifted to be resonant within a detuning $\Delta$ with the $6P_{3/2}$ intermediate state, which decays to the ground state with a decay constant $1/\tau_{6P}$. After de-excitation, an electric field ionizes all Rydberg states with principal quantum number $n\gtrsim 55$.  }
\label{Fig.1}
\end{center}
\end{figure}

In order to excite and de-excite Rydberg atoms \cite{Mudrich:2005, Day:2008, Karlewski:2015}, we shift the frequency of the $1013 \,\mathrm{nm}$  laser using an AOM in double-pass configuration and with a radio-frequency driving that allows us to switch between excitation and de-excitation (with Rabi frequency $\Omega_{1013}$) within a few hundred $ \,\mathrm{ns}$. This procedure is illustrated in Fig. \ref{Fig.1}. We checked that over the range of frequencies used, the position of the beam varies by less than  $10 \,\mathrm{\mu m}$ relative to the MOT. Also, the amplitude of the radio-frequency driving is calibrated to maintain a constant intensity of the $1013 \,\mathrm{nm}$ beam for different driving frequencies.

The protocol for our experiments is as follows. Initially, an excitation pulse of duration $t_{ex}$ is applied with both lasers, where the two-photon excitation is detuned by $\Delta_{ex}$ from resonance (the MOT beams are switched off during the entire excitation and de-excitation sequence). After a variable dark time $t_{dark}$, during which both lasers are switched off, only the $1013 \,\mathrm{nm}$  laser is switched on for $t_{deex}$, with the AOM frequency set to a value that shifts the frequency of that laser to be resonant with the transition $70S_{1/2}-6P_{3/2}$ to within a detuning $\Delta$. The $6P_{3/2}$ state has a lifetime $\tau_{6P} \approx 120  \,\mathrm{ns}$. Finally, $300 \,\mathrm{ns}$ after the de-excitation pulse an electric field is switched on that field ionizes the Rydberg atoms (in practice all Rydberg atoms with $n \gtrsim 55$) and accelerates the ions towards a channeltron, where they are detected with an overall efficiency $\eta \approx 0.4$, details are described in \cite{Viteau:2013, Malossi:2014}. The experimental cycle is repeated $100$ times, and the mean number of excitations $\langle N\rangle$ as well as the variance are calculated. The values we report in this work have been corrected for the detection efficiency $\eta$ \cite{Malossi:2014}.

Numerical simulations of our experiments are performed using model system that capture the essential features of our experiments. For the blockade regime, defined below, a two-dimensional arrangement of $300$ atoms with fixed positions separated by $a=r_b/10$ along the x-direction and random positions in a range between $-1.5\, r_b$ and $+ 1.5 \,r_b$ are used in the perpendicular direction. Excitation and de-excitation in the incoherent regime are simulated using a Kinetic Monte Carlo algorithm \cite{Chotia:2008} (on the short timescales for excitation and de-excitation we neglect atomic motion). In the simulations of the de-excitation process, we put $\Omega_{1013} =2 \pi \times 1\,\mathrm{MHz}$ and adapted the simulation parameter $t_{deex}$ to obtain quantitative agreement with the experimental results. For the facilitation regime, a simplified simulation is used, in which $100$ atoms are initially placed at a fixed spacing $r_{fac}$ along the x-axis, and initially $20$ Rydberg excitations are created at odd positions. A further $20$ excitations are then added at random even positions, resulting in a 1D chain with defects (which reflects the fact that the facilitation process is unlikely to produce perfect 1D chains). Thermal motion before the de-excitation pulse is simulated by averaging over initial configurations with $r_{fac}$ increased to $\alpha r_{fac}$ with $\alpha$ between $1$ and $1.3$ according to a Boltzmann distribution reflecting the thermal velocity distribution of the atoms. Finally, de-excitation is simulated as in the blockade regime.

\begin{figure}[htbp]
\begin{center}
\includegraphics[width=8cm]{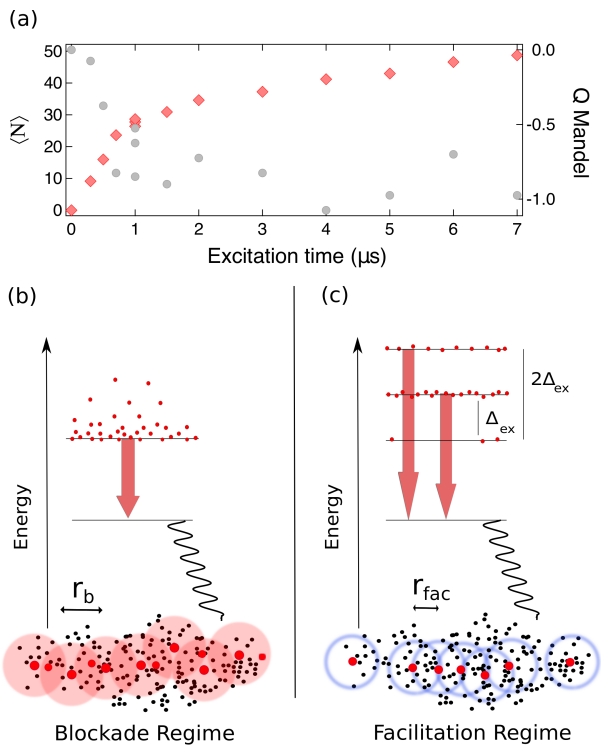}
\caption{The distinction between the non-interacting and the interacting regimes for resonant excitation is shown in (a), where the mean number of excitations $\langle N\rangle$ (red diamonds) and the Mandel $Q$-parameter (grey circles) are plotted as a function of excitation time. The crossover between the two regimes is signalled by a decrease in the slope of $\langle N\rangle$  as a function of $t_{ex}$ and by a $Q$ approaching $-1$. The corresponding energy distribution (red dots) of the Rydberg atoms during the de-excitation process in the interacting regime is shown in (b), where excitations at distances smaller than $r_b$ (blockade volume represented by transparent red discs) are present. In the latter case, the de-excitation resonances are shifted in energy with respect to the non-interacting case for which they are located at $\Delta=0$. In the facilitation regime (c) Rydberg atoms are created at the facilitation distance $r_{fac}$ by off-resonant excitation (blue rings represent the facilitation shells), leading to de-excitation resonances at $\Delta=\Delta_{ex}$ and $\Delta= 2 \Delta_{ex}$ for atoms at the edges and inside the (quasi) one-dimensional chain, respectively. }
\label{Fig.2}
\end{center}
\end{figure}

As one of the aims of this work is to characterize the effects of the van der Waals interaction between Rydberg excitations on the de-excitation dynamics, we first define three different interaction regimes. For resonant excitation ($\Delta_{ex} = 0$), the non-interacting regime is realized by limiting the number of excitations $N_{in}$ such that the mean distance $d$ between Rydberg atoms is larger than the blockade radius $r_b$. The interacting (or blockade) regime, on the other hand, is reached when $d \ll r_b$. Empirically we identify the two regimes by analyzing the dynamics of $\langle N\rangle$ and the fluctuations during the excitation pulse (as in our previous works \cite{Viteau:2012,Malossi:2014}. A typical example of that dynamics is shown in Fig. \ref{Fig.2}(a), where $\langle N\rangle$ and the Mandel $Q$ parameter \cite{Mandel:1982} $Q =(\langle N^2\rangle-\langle N\rangle^2)/\langle N\rangle -1$ are plotted as a function of $t_{ex}$. The dynamics slows down appreciably when $\langle N\rangle \approx 30$ as a consequence of the dipole blockade, and $Q$ drops below zero, which signals sub-Poissonian counting statistics. Finally, the facilitation regime corresponds to off-resonant excitation ($\Delta_{ex} > 0$) and is characterized by a positive value of $Q$ \cite{Viteau:2012}. 

\begin{figure}[htbp]
\begin{center}
\includegraphics[width=6cm]{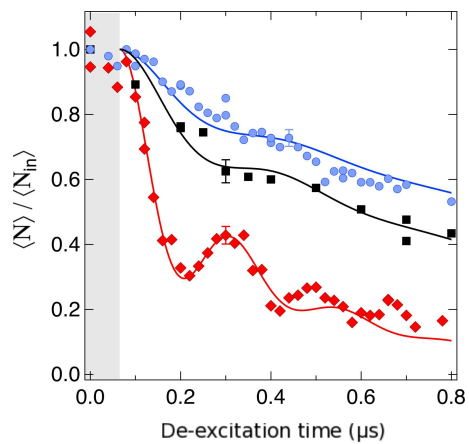}
\caption{Normalized population of the $70S_{1/2}$ Rydberg state as a function of the de-excitation time $t_{deex}$ for different de-excitation Rabi frequencies $\Omega_{1013}$ . Damped Rabi oscillations are visible for $\Omega_{1013} \approx 2 \pi \times4\,\mathrm{MHz}$ (red diamonds), whereas for smaller $\Omega_{1013}$  ($2 \pi \times 2\,\mathrm{MHz}$, black squares, and $2 \pi \times1.4\,\mathrm{MHz}$, blue circles) they are strongly reduced. The experimental data are well reproduced by the simple damped two-level model described in the main text (solid lines). Note that the numerical simulations are shifted by $60\,\mathrm{ns}$ to take into account the finite rise time of the acousto-optic modulator, leading to a nonlinear relationship between the pulse time and pulse area during the first $\approx 100 \,\mathrm{ns}$.}
\label{Fig.3}
\end{center}
\end{figure}

We first illustrate the de-excitation technique in the non-interacting regime. Fig. \ref{Fig.3} shows the fraction $\langle N\rangle /\langle N_{in}\rangle$ of Rydberg atoms remaining after the excitation pulse as a function of $t_{deex}$ for different values of $\Omega_{1013}$. For large values of $\Omega_{1013}$ around $2\pi \times 4  \,\mathrm{MHz}$ (which is larger than the frequency associated with decay from the $6 P_{3/2}$ state, 1/$\tau_{6P} \approx 2 \pi \times 1.3 \,\mathrm{MHz}$), the dynamics shows signs of residual coherent oscillations, whereas  for $\Omega_{1013}$ below $ 2 \pi \times 2 \,\mathrm{MHz}$ those oscillations are strongly damped and the de-excitation dynamics can, to a good approximation, be described by an exponential decay. Fig. \ref{Fig.3}  also shows the results of a numerical integration of a simple (coherent) two-level system with loss terms from the Rydberg state and the $6 P_{3/2}$  state, with respective decay rates 1/$\tau_{70S}$ and 1/$\tau_{6P}$ (the lifetime $\tau_{70S} \approx 150  \,\mathrm{\mu s}$ is much larger than all the other timescales in our problem and can be neglected for the purposes of this work). In the remainder of this work $\Omega_{1013} < 2 \pi \times 2 \,\mathrm{MHz}$, so that, similarly to the excitation process discussed above, the single-atom de-excitation dynamics can always be described by a rate equation with $\Gamma= \frac{\Omega_{1013}^2}{2 \gamma}\cdot\frac{1}{|1+(\Delta/\gamma)^2|}$. 

\begin{figure}[htbp]
\begin{center}
\includegraphics[width=9cm]{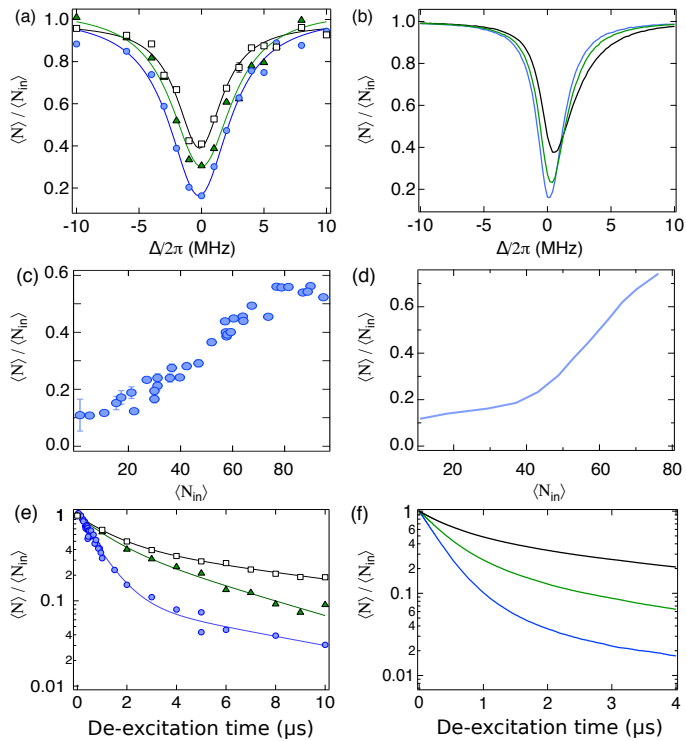}
\caption{ De-excitation process following resonant excitation for different initial mean numbers $\langle N_{in}\rangle$:  25 (blues circles), 34 (green triangles) and 50 (white squares). The different values of $\langle N_{in}\rangle$ (ranging from the non-interacting to the interacting regime) are obtained by varying $t_{ex}$ between $0.5  \,\mathrm{\mu s}$ and $5  \,\mathrm{\mu s}$. In (a), the remaining fraction of Rydberg atoms $\langle N \rangle /\langle N_{in}\rangle$ is plotted as a function of the de-excitation detuning $\Delta$. Here, $t_{dark} = 0.5  \,\mathrm{\mu s}$ and $t_{deex} = 2   \,\mathrm{\mu s}$. The solid lines are Lorentzian fits to guide the eye. The expected shift in the de-excitation detuning is visible mainly as an increase in the remaining fraction at $\Delta=0$, shown systematically in (c). Here, the remaining fraction after a de-excitation pulse of duration $t_{deex} = 1 \,\mathrm{\mu s}$ is plotted as a function of $\langle N_{in}\rangle$ . In (e), the de-excitation dynamics is shown for $\Delta=0$. In (a), (c), and (e), error bars are one standard deviation of the mean. Panels (b), (d) and (f) show the results of the corresponding numerical simulations (see main text). In (b) and (d) we adjust the parameter $t_{deex}$ equal to $0.5\,\mathrm{\mu s}$.}
\label{Fig.4}
\end{center}
\end{figure}

We now proceed to systematically study the de-excitation dynamics in the different interaction regimes. As shown above, in the non-interacting (and incoherent) regime the dynamics can be described by a rate equation, leading to an exponential decrease of $\langle N\rangle $ with $t_{deex} $. For resonant excitation one, therefore, expects to see a minimum in the remaining fraction of Rydberg excitations after the de-excitation pulse as a function of the detuning $\Delta$ for $\Delta=0$. This is confirmed in Fig. \ref{Fig.4}(a), where for an initial $\langle N_{in}\rangle = 20$, $\langle N\rangle /\langle N_{in}\rangle$ is plotted as a function of $\Delta$ for a fixed $t_{deex} = 2  \,\mathrm{\mu s}$ and $\Omega_{1013} = 2\pi \times 1 \,\mathrm{MHz}$. When $\langle N_{in}\rangle$ is increased to $50$, corresponding to the interacting regime, the remaining fraction at $\Delta=0$ also increases, indicating that the interactions shift the Rydberg levels and hence the de-excitation laser is no longer resonant. This effect is shown more systematically in Fig. \ref{Fig.4}(c), where the remaining fraction of Rydberg excitations at $\Delta=0$ is plotted as a function of $\langle N_{in}\rangle$. Between the non-interacting regime ($\langle N_{in}\rangle \approx 2$, corresponding to an interatomic distance $d \approx 70 \,\mathrm{\mu m} > r_b$) and the strongly interacting regime ($\langle N_{in}\rangle \approx 80$, for which $d \approx 2 \,\mathrm{\mu m} < r_b$), the remaining fraction increases from $0.1$ to $0.6$. This crossover from the non-interacting to the interacting regime is also visible in the de-excitation dynamics. Fig. \ref{Fig.4}(e) shows the remaining fraction as a function of $t_{deex}$ for different values of $\langle N_{in}\rangle$. The de-excitation rate decreases appreciably (by up to a factor 6) as $\langle N_{in}\rangle$ is increased. The effect of the van der Waals interactions is also reflected in the fact that the dynamics of the remaining fraction does not follow a simple exponential decay. Rather, the rate of the exponential decay decreases as $t_{deex}$ is increased. We interpret this as a consequence of the spread of inter-atomic distances between the excited atoms, which means that Rydberg atoms with more distant neighbours are de-excited faster, whereas those interacting more strongly with their closer neighbours exhibit reduced de-excitation rates. In Fig. \ref{Fig.4}(b,d,f) we show the results of numerical simulations, which agree well with our experiments and reproduce the salient features of our observations.

\begin{figure}[htbp]
\begin{center}
\includegraphics[width=14cm]{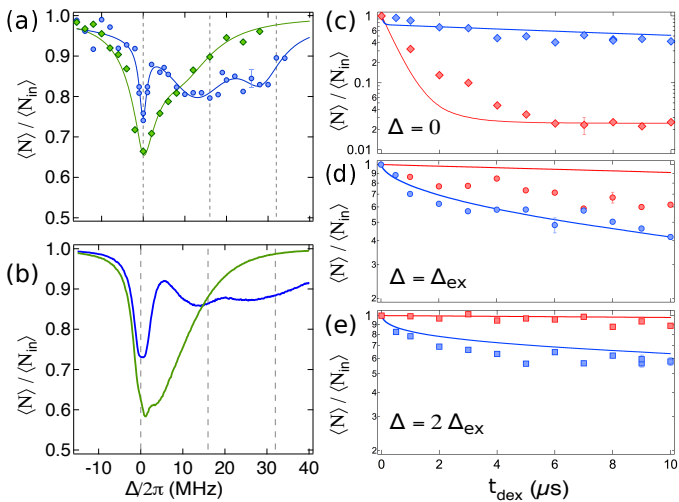}
\caption{De-excitation process following off-resonant excitation in the facilitation regime. In (a) the remaining fraction  $\langle N \rangle /\langle N_{in}\rangle$  after excitation of $\langle N_{in}\rangle \approx 20$ excitations ($t_{ex} = 5 \,\mathrm{\mu s}$) at $\Delta_{ex} = 2\pi \times 16  \,\mathrm{MHz}$ is plotted as a function of $\Delta$. The blue circles correspond to de-excitation ($t_{deex} = 2  \,\mathrm{\mu s}$) after a dark time $t_{dark} = 0.5  \,\mathrm{\mu s}$, whereas the green diamonds are obtained for $t_{dark} = 5  \,\mathrm{\mu s}$. For comparison, a Lorentzian fit to data obtained after resonant excitation ($t_{ex} = 0.5 \,\mathrm{\mu s}$) is also shown (dashed line). The solid lines are triple-Lorentzian fits to guide the eye. The results of the corresponding numerical simulations are shown in (b), where the solid lines corresponds to de-excitation of an artificial state with $70$ atoms placed on a line at distance of $r_{fac}$ ( blue line) or $\alpha r_{fac}$ ( green line) as described in the main text. Plots (c)-(e) illustrate the experimental data (symbols) and the simulation results (solid lines ) of the de-excitation dynamics ($t_{dark} = 0.5  \,\mathrm{\mu s}$) at $\Delta=0$ (c), $\Delta=\Delta_{ex}$ (d) and $\Delta= 2\times\Delta_{ex}$ (e) following resonant excitation (red symbols) and off-resonant excitation at $\Delta_{ex} = 2\pi \times 16 \,\mathrm{MHz}$ (blue symbols). In the experiments, the same values $\langle N_{in}\rangle \approx 20$ are obtained in both cases for $t_{ex}=1  \,\mathrm{\mu s}$  by adjusting $\Omega$. }
\label{Fig.5}
\end{center}
\end{figure}

In the facilitation regime ( i.e., for off-resonant excitation) the role of the interactions is substantially different compared to the cases described above. Here, at a distance $r_{fac}$ from an excited atom the interactions facilitate further excitations by shifting the off-resonant laser driving with detuning $\Delta_{ex}$ into resonance. In the quasi one-dimensional geometry considered in this work, this process results in a chain of excitations at a fixed spacing $r_{fac}$. Atoms at the edges of this chain have a single neighbour and hence an interaction energy $\hbar \Delta_{ex}$, whereas atoms inside the chain have two neighbours and a resulting energy shift of $2\hbar \Delta_{ex}$. The de-excitation resonances for those two classes of atoms should, therefore, be centred around $\Delta=\Delta_{ex}$ and $\Delta=2\Delta_{ex}$, respectively. Furthermore, due to the residual thermal motion of the atoms (and also the van der Waals repulsion \cite{Thaicharoen:2015, Teixeira:2015, Faoro:2016}) the distances between the atoms will increase over time, so that eventually each atom will have a de-excitation resonance at $\Delta=0$ as the interactions decrease. 
We test the above picture by off-resonantly exciting around $\langle N_{in}\rangle = 20$ atoms at $\Delta_{ex}=2\pi \times 16   \,\mathrm{MHz}$ using a $5  \,\mathrm{\mu s}$ excitation pulse. As the $70S_{1/2}$ Rydberg state used here interact repulsively, the facilitation condition correspond to a positive detuning.  A de-excitation pulse of duration $t_{deex}=2  \,\mathrm{\mu s}$ follows after two different values of a dark time: $t_{dark} = 0.5  \,\mathrm{\mu s}$ and $t_{dark} = 5  \,\mathrm{\mu s}$. In Fig. \ref{Fig.5}(a), for $t_{dark} = 0.5  \,\mathrm{\mu s}$ three de-excitation resonances can be seen, corresponding to atoms with two neighbours at distance $r_{fac}$ ($\Delta=2\Delta_{ex}$), with one neighbour ($\Delta=\Delta_{ex}$), and without any neighbours ($\Delta=0$ ). The latter class of atoms corresponds to single off-resonantly excited Rydberg atoms that did not lead to further facilitation events, or else to atoms whose neighbours at $r_{fac}$ have already moved sufficiently so as to reduce the interaction energy effectively to zero (due to the $1/r^6$ dependence of the van der Waals interaction, a 50$\%$ increase in the interatomic distance leads to a reduction in the interaction energy by one order of magnitude).
When $t_{dark}$ is increased to $5  \,\mathrm{\mu s}$, the effects of thermal motion are clearly visible. The de-excitation resonance at $\Delta=0$  is now more pronounced, whereas those at $\Delta=\Delta_{ex}$ and $\Delta=2\Delta_{ex}$ are substantially reduced. This observation agrees with the fact that for our MOT temperatures the atoms move, on average, $0.6 \,\mathrm{\mu m}$ in $5\,\mathrm{\mu s}$, which leads to a reduction of the interaction energy between excited atoms to around $50\%$ of its initial value. 

The qualitative behaviour of the de-excitation spectrum agrees well with our numerical simulation shown in Fig.5 (b), for $0 \mathrm{\mu s}$ (blue line) and $5 \mathrm{\mu s}$  (green line) between excitation and de-excitation.

The difference between the de-excitation dynamics after resonant and off-resonant excitation, respectively, is summarized in Fig. \ref{Fig.5}(c-e). Here, for the two excitation regimes the de-excitation dynamics is shown for three different values of $\Delta$: $\Delta=0$, $\Delta=\Delta_{ex}$ and $\Delta=2\Delta_{ex}$. Whereas de-excitation at $\Delta=0$ is appreciable if the atoms were initially excited on resonance, the de-excitation dynamics for $\Delta=\Delta_{ex}$ and $\Delta=2\Delta_{ex}$ is essentially negligible in that case. By contrast, for excitation in the facilitation regime, the de-excitation dynamics for $\Delta=\Delta_{ex}$ and $\Delta=2\Delta_{ex}$ is clearly visible. Good qualitative agreement is obtained with our numerical simulations. To account for blackbody-induced migration to other (no longer de-excitable) Rydberg states, we impose a lower limit of the de-excitable fraction ($2.5\%$) that best fits our experimental data. This value agrees with the results of experiments currently under way in our laboratory, in which we measured the de-excitable fraction as a function of time.

In summary, we have investigated de-excitation from Rydberg states via a fast decaying intermediate state in a variety of conditions. In particular, we find that the effect of the van der Waals interactions between Rydberg atoms is clearly visible both in the de-excitation spectrum (as a function of $\Delta$) and in the de-excitation dynamics at fixed $\Delta$. For off-resonant excitation in the facilitation regime, we see three distinct de-excitation resonances which reflect the different interaction energies due to the spatial arrangement of the atoms. In future experiments, this feature might be used to obtain information about the spatial ordering in Rydberg clusters or to manipulate such clusters by selectively de-exciting Rydberg atoms satisfying a particular energetic condition. We also note that the clear evidence shown here for the effect of interactions on the de-excitation dynamics has implications for state-selective detection techniques using depumping to the ground state in order to count Rydberg excitations or reveal their position. In such schemes, interaction effects have to be taken into account or compensated in order to avoid systematic errors. The same reasoning applies to controlled dissipation techniques, where decay from a Rydberg state is accelerated by de-excitation via a fast decaying state.

The authors acknowledge support by the European Union H2020 FET Proactive project RySQ (grant N. 640378) and by the EU Marie Curie ITN COHERENCE (Project number 265031).

\section*{References}
\bibliographystyle{apsrmp}

\begin{thebibliography}{0}
\expandafter\ifx\csname natexlab\endcsname\relax\def\natexlab#1{#1}\fi
\expandafter\ifx\csname bibnamefont\endcsname\relax
  \def\bibnamefont#1{#1}\fi
\expandafter\ifx\csname bibfnamefont\endcsname\relax
  \def\bibfnamefont#1{#1}\fi
\expandafter\ifx\csname citenamefont\endcsname\relax
  \def\citenamefont#1{#1}\fi
\expandafter\ifx\csname url\endcsname\relax
  \def\url#1{\texttt{#1}}\fi
\expandafter\ifx\csname urlprefix\endcsname\relax\def\urlprefix{URL }\fi
\providecommand{\bibinfo}[2]{#2}
\providecommand{\eprint}[2][]{\url{#2}}

\end{thebibliography}


\begin{thebibliography}{99}
\bibitem{Ga�tan:2009} A. Ga\"etan, Y. Miroshnychenko, T. Wilk, A. Chotia, M. Viteau, D. Comparat, P. Pillet, A. Browaeys, and P. Grangier, Nat. Phys. 5, 115 (2009).
\bibitem{G�rttner:2014} M. G\"arttner, S. Whitlock, D. W. Sch\"onleber, and J\"org Evers Phys. Rev. Lett. 113, 233002 (2014).
\bibitem {Schau�:2012} P. Schau{\ss}, M. Cheneau, M. Endres, T. Fukuhara, S. Hild, A. Omran, T. Pohl, C. Gro{\ss}, S. Kuhr, and I. Bloch, Nature 491, 87 (2012).
\bibitem{Urban:2009} E. Urban, T. A. Johnson, T. Henage, L. Isenhower, D. D. Yavuz, T. G. Walker, and M. Saffman, Nat. Phys. 5, 110 (2009).
\bibitem{Saffman:2013} M. Saffman, T. G. Walker, and K. M\o lmer Rev. Mod. Phys. 82, 2313 (2010).
\bibitem{Lesanovsky:2013} I. Lesanovsky and J. P. Garrahan, Phys. Rev. Lett. 111, 215305 (2013).
\bibitem{Urvoy:2015} A. Urvoy, F. Ripka, I. Lesanovsky, D. Booth, J. P. Shaffer, T. Pfau, and R. L\"ow, Phys. Rev. Lett. 114, 203002 (2015).
\bibitem{Valado:2016} M.M. Valado, C. Simonelli, M. D. Hoogerland, I. Lesanovsky,  J.P. Garrahan, E. Arimondo, D. Ciampini, and O. Morsch, Phys. Rev. A 93, 040701(R) (2016).
\bibitem{Marcuzzi:2016} M. Marcuzzi, M. Buchhold, S. Diehl, and I. Lesanovsky, Phys. Rev. Lett. 116, 245701 (2016)
\bibitem{Dudin:2012} Y. O. Dudin and A. Kuzmich, Science 336, 887 (2012).
\bibitem{Simonelli:2016} C. Simonelli, M. M. Valado, G. Masella, L. Asteria, E. Arimondo, D. Ciampini and O. Morsch, J. Phys. B: At. Mol. Opt. Phys. 49, 154002 (2016) 
\bibitem{Lesanovsky:2014} I. Lesanovsky and J. P. Garrahan, Phys. Rev. A 90, 011603(R) (2014).
\bibitem{Mudrich:2005} M. Mudrich, N. Zahzam, T. Vogt, D. Comparat, and P. Pillet, Phys. Rev. Lett. 95, 233002 (2005).
\bibitem{Day:2008} J. O. Day, E. Brekke, and T. G. Walker, Phys. Rev. A 77, 052712  (2008).
\bibitem{Karlewski:2015} F. Karlewski, M. Mack, J. Grimmel, N. S\'andor, and J. Fort\'agh, Phys. Rev. A 91, 043422 (2015).
\bibitem{Viteau:2013} M. Viteau, M. Bason, J. Radogostowicz, N. Malossi, O. Morsch, D. Ciampini, and E. Arimondo, Laser Phys. 23, 015502 (2013).
\bibitem{Malossi:2014} N. Malossi, M. M. Valado, S. Scotto, P. Huillery, P. Pillet, D. Ciampini, E. Arimondo, and O. Morsch, Phys. Rev. Lett. 113, 023006 (2014).
\bibitem{Chotia:2008} A. Chotia, M. Viteau,T. Vogt, D. Comparat, and P. Pillet, New Journal of Physics 10, (2008).
\bibitem{Mandel:1982} L. Mandel, Phys. Rev. Lett. 49, 136 (1982)
\bibitem{Viteau:2012} M. Viteau, P. Huillery, M. G. Bason, N. Malossi, D. Ciampini, O. Morsch, E. Arimondo, D. Comparat, and P. Pillet, Phys. Rev. Lett. 109, 053002 (2012).
\bibitem{Thaicharoen:2015} N. Thaicharoen, A. Schwarzkopf, and G. Raithel, Phys. Rev. A 92, 040701(R) (2015).
\bibitem{Teixeira:2015} R. C. Teixeira, C. Hermann-Avigliano, T. L. Nguyen, T. Cantat-Moltrecht, J. M. Raimond, S. Haroche, S. Gleyzes, and M. Brune, Phys. Rev. Lett. 115, 013001 (2015).
\bibitem{Faoro:2016} R. Faoro, C. Simonelli, M. Archimi, G. Masella, M. M. Valado, E. Arimondo, R. Mannella, D. Ciampini, and O. Morsch, Phys. Rev. A, 93, 030701(R) (2016).






\end{thebibliography}




\end{document}